\documentclass[runningheads]{llncs}
\usepackage[T1]{fontenc}
\usepackage{graphicx}
\usepackage{subcaption}
\usepackage{booktabs}
\usepackage{amsmath}
\usepackage{amssymb}
\usepackage{multirow}
\usepackage{xcolor}
\begin{document}
\title{X-GAN: A Generative AI-Powered Unsupervised Model for Main Vessel Segmentation of Glaucoma Screening}
%
\author{Cheng Huang\inst{1,3} \and
Weizheng Xie\inst{1} \and Tsengdar Lee\inst{2}
 \and Karanjit Kooner\inst{3}$^{,\dag}$ \and Jui-Kai Wang\inst{3} \and Ning Zhang\inst{4} \and Jia Zhang\inst{1}$^{,\dag}$
}
\authorrunning{F. Author et al.}
%
\institute{Southern Methodist University \and
National Aeronautics and Space Administration
\and
University of Texas Southwestern Medical Center
\and
Northeastern University
\\
$\dag$ Corresponding Author\\
\email{jiazhang@smu.edu}}

\maketitle              
\begin{abstract}
Structural changes in main retinal blood vessels serve as critical biomarkers for the onset and progression of glaucoma. Identifying these vessels is vital for vascular modeling yet highly challenging. This paper proposes X-GAN, a generative AI-powered unsupervised segmentation model designed for extracting main blood vessels from Optical Coherence Tomography Angiography (OCTA) images. The process begins with the Space Colonization Algorithm (SCA) to rapidly generate a skeleton of vessels, featuring their radii. By synergistically integrating the generative adversarial network (GAN) with biostatistical modeling of vessel radii, X-GAN enables a fast reconstruction of both 2D and 3D representations of the vessels. Based on this reconstruction, X-GAN achieves nearly 100\% segmentation accuracy without relying on labeled data or high-performance computing resources. Experimental results confirm X-GAN’s superiority in evaluating main vessel segmentation compared to existing deep learning models.

\end{abstract}
\section{Introduction}

Glaucoma is a leading cause of irreversible blindness, often progressing silently until significant vision loss occurs \cite{kooner2024meeting,fairclip,fairdomain,GDP,kooner2022glaucoma}. Recent research highlights choroidal microvasculature dropout as a potential biomarker for disease progression \cite{kooner2022glaucoma}. Optical Coherence Tomography Angiography (OCTA) enables visualization of these vascular changes, but accurately assessing choroidal vessel density remains a challenging task \cite{octa-500,glalstm,glaboost,GAN-Unet}.

Existing medical imaging analytics in glaucoma research predominantly depend on computer vision techniques \cite{octa-500-32d,GAN-Unet,skin1,fairseg,fairclip,fairdomain,Fairness,GDP,mask2,FairVision,lag1,cnn-2,cnn-1,unet-gla}. In recent years, supervised learning segmentation models (SLSMs) \cite{octa-500-32d,octa-500-gen,UNet++,MedSAM,mask,unet,aunet,mask-1,CauSSL,UniverSeg,S2VNet,lag1,cnn-2,cnn-1,unet-gla} have become particularly prominent in this field. However, SLSMs may not be well-suited for detecting choroidal vessels in OCTA images. They rely heavily on labeled data, but manually annotating choroidal blood vessels in OCTA images is extremely challenging with current devices \cite{octa-500,ROSE}. OCTA images offer high-resolution, depth-resolved visualization of retinal and choroidal microvasculature without the need of contrast dye, making them more effective than traditional fundus imaging for detecting vascular abnormalities \cite{octa-500,ROSE}. However, segmenting choroidal vessels is difficult due to irregular patterns, intersecting pathways, dense capillary networks \cite{kooner2022glaucoma}, and complex capillary structures. Moreover, the presence of major blood vessels, which must be excluded, further complicates the labeling process \cite{kooner2024meeting,kooner2022glaucoma}. 

Glaucoma researchers have revealed that retinal blood vessels follow biostatistical relationships, meaning that vessel radius variations along branching structures \cite{campbell2017detailed,GAN-Unet}. raising the question of whether segmentation can bypass pixel-based mapping and leverage vessel radius as a key marker. To address this, we propose X-GAN, an unsupervised model that integrates GANs \cite{GAN} with retinal vessel biostatistics for precise main vessel segmentation from OCTA images, eliminating the need for labeled data or extensive training. 

In summary, the contributions of this paper are three-fold:

\begin{itemize}
\item We introduce X-GAN, a generative AI-powered unsupervised segmentation model tailored for extracting main blood vessels from OCTA images without annotations. 
\item We eliminate the reliance on synthetic label supervision mechanisms by introducing a novel segmentation strategy based on vessel radius thresholds and depth-first search (DFS), enabling accurate main vessel extraction without any pixel-level annotations.
\item Experimental results of X-GAN for retinal main vessel segmentation have demonstrate that X-GAN outperformed state-of-the-art (SOTA) current models, achieving nearly 100\% segmentation accuracy. 
\end{itemize}

\section{Related Work}

\subsection{Medical Imaging in Glaucoma}
Many datasets of glaucoma focus on computer vision tasks \cite{octa-500,ROSE,fairseg,fairclip,fairdomain,Fairness,GDP,FairVision,lag1,dataset-DRIVE,dataset-ORIGA,dataset-Drishti-GS}, especially in medical image processing. For inner retinal blood vessels, datasets like LAG \cite{lag1}, OCTA-500 \cite{octa-500} and ROSE \cite{ROSE}, provide various-sizeimages (e.g., 3mm, 6mm). However, these images represent only a small portion of the total dataset. Some datasets \cite{fairseg,fairclip,fairdomain,Fairness,GDP,FairVision,lag1,dataset-DRIVE,dataset-ORIGA,dataset-Drishti-GS} offer fundus images, and the capillaries are nearly indistinguishable to the naked eye. There is a shortage of high-quality, intuitive data, such as detailed images of inner 3mm$\times$3mm retinal vessels, and existing datasets often lack sufficient clarity. Consequently, few studies focus on structural changes in major retinal blood vessels in glaucoma.

\subsection{AI for Glaucoma}

Even with accessible data, labeling remains challenging due to annotation costs and inter-observer variability, introducing bias and hindering model performance \cite{fairseg,fairclip,fairdomain,Fairness,GDP,FairVision,unet-gla}. Generative AI can alleviate data scarcity and labeling constraints, though human oversight is often required \cite{gai}. GANs \cite{gan-d-1,GAN-m1,ct-gan,GAN-cance,GAN,CycleGAN}, as unsupervised models, address both issues by generating realistic samples without manual labels. In this framework, segmentation is typically achieved by coupling the GAN with a dedicated segmentor $S$, which can be implemented using pre-trained CNNs \cite{lag1,cnn-2,cnn-1}, U-Net variants \cite{UNet++,unet,aunet,unet-gla}, ViT \cite{vit}, Mask R-CNN \cite{mask}, or MedSAM \cite{MedSAM}, enabling end-to-end vessel synthesis and segmentation.

\subsection{Motivation}

In traditional models such as GAN+SLSM \cite{GAN-Unet,gan-d-1,DO-GAN,ct-gan,TR-GAN,GAN-cance} and standalone SLSMs \cite{octa-500-32d,octa-500-gen,UNet++,MedSAM,mask,unet,aunet,CauSSL,UniverSeg,S2VNet,lag1,cnn-2,cnn-1,unet-gla}, pixel-level segmentations are prone to inter-observer variability \cite{octa-500,octa-500-32d,GDP}. However, in OCTA, retinal vessels have consistent visual traits, uniform grayscale gradients and path-like shapes, making them suitable for coordinate-based modeling with associated widths.

We adopt the Space Colonization Algorithm (SCA) \cite{sca} to represent vessels as structured graphs, enhancing topological coherence and connectivity. To bridge the domain gap between initial vessel maps and real OCTA data, we apply a CycleGAN-based refinement module \cite{CycleGAN}, aligning generated structures with realistic image characteristics while preserving topology. Unlike prior methods that segment post-rendered images, we introduce a DFS-based segmentation approach \cite{tarjan1972dfs} applied directly to graph-structured (coordinates, radius) data. This intercepts the generation process to ensure anatomical fidelity, governed by a radius threshold $R_{min}$ derived from biostatistical vessel metrics \cite{kooner2024meeting,bohm2023methods,campbell2017detailed,niemeijer2010automatic}.

\section{Methodology}

\begin{figure*}[ht]
  \centering
   \centerline{\includegraphics[width=0.8\columnwidth]{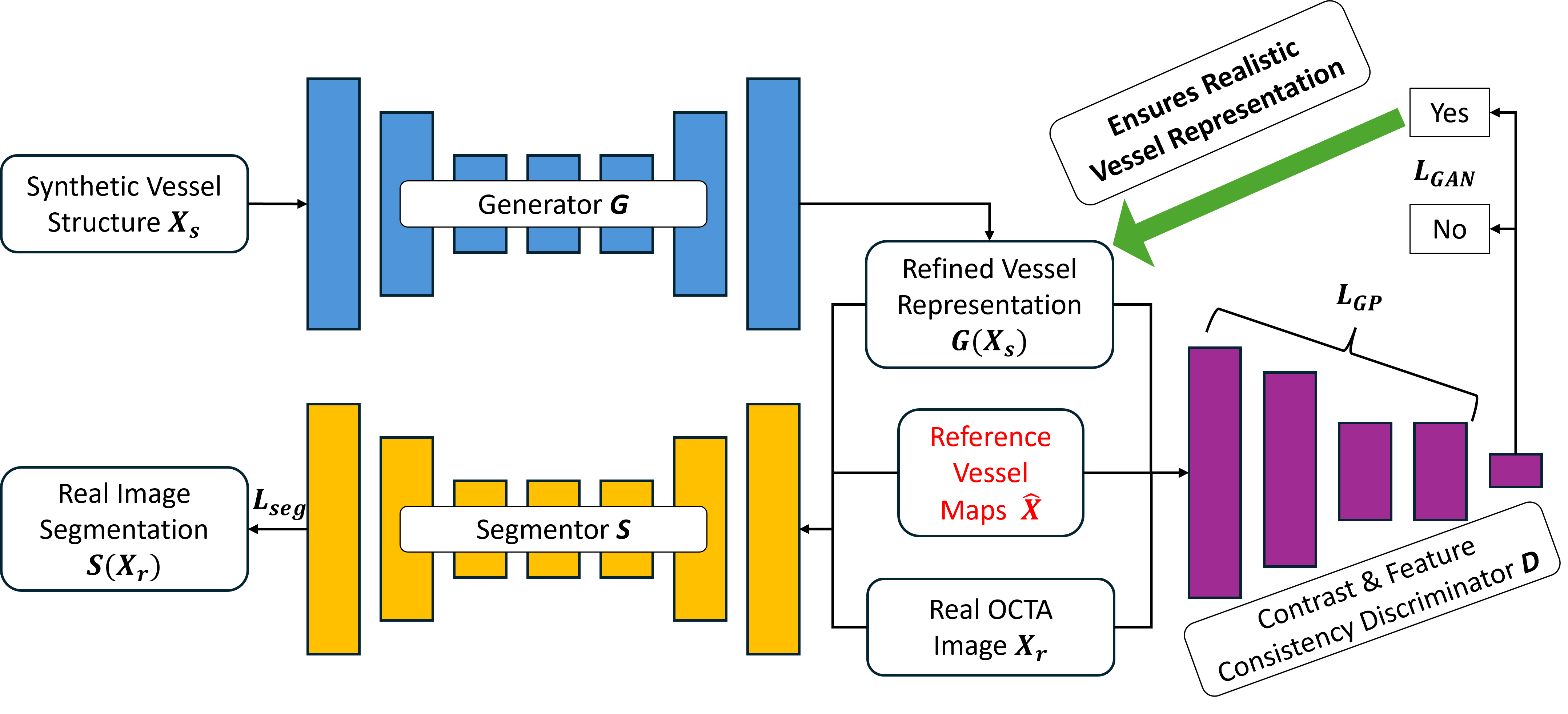}}
    \caption{The architecture of X-GAN where red indicates that X-GAN does not utilize these things in Segmentor $S$, whereas GAN+SLSMs do.}
    \label{s1}
\end{figure*}

The complete framework of X-GAN is illustrated in Fig.~\ref{s1} and it consists of two parts: Vessel Structure Refinement Module Module and Segmentor.

\subsection{Vessel Structure Refinement Module}

Rather than generating pixel-wise vessel segmentations, we formulates the vascular network as a structured graph representation, ensuring topological consistency and robust connectivity. The graph consists of the following components, shown in Equation~\ref{sca}:
\begin{equation}
  V = \left\{ (x_i, y_i, z_i | r_i) \right\}_{i=1}^{N}
  \label{sca}
\end{equation}
where ($x_{i}$, $y_{i}$, $z_i$) are the spatial coordinates of the vessel centerline points, $r_{i}$ represents the local vessel radius and $N$ means the number of nodes. This structured representation is initially generated using the SCA \cite{sca,sca-1,sca-2}, which models vessel growth based on attraction points and bifurcation principles. However, the initial vessel maps exhibit domain discrepancies in contrast and noise characteristics compared to real OCTA images. To mitigate this gap, we employ a tuned GAN based on CycleGAN \cite{CycleGAN} as a structural refinement module, aligning vessel representations with real OCTA distributions while preserving vascular topology.

Given a vessel structure $X_{s}$ via SCA, our generator $G$ learns to transform it into a realistic OCTA-like representation $X_r'$, expressed as $G(X_s)$, while $X_r'$ undergoes style adaptation to align with the contrast and noise characteristics of real OCTA images $X_r$. To enhance structural consistency, we modify the original CycleGAN architecture by removing the inverse generator and incorporating a segmentation consistency loss via the segmentor $S$. This modification prevents CycleGAN from altering the vessel topology, ensuring that vessel segmentation remains biologically meaningful, as illustrated in Equation~\ref{lu}:
\begin{equation}
    L_{\text{seg}} = \mathbb{E}_{X_r} \left[ \| S(X_r') - S(X_r) \|_1 \right]
   \label{lu}
\end{equation}
where $S(X)$ extracts vessel masks from OCTA images, enforcing structural consistency during transformation. This constraint ensures that CycleGAN adapts contrast while strictly preserving vessel topology. Consequently, our total loss function is formulated as follows:

\begin{align}
    L_{\text{GAN}} &= \mathbb{E}_{X_r} [D(X_r)] - \mathbb{E}_{X_s} [D(G(X_s))] 
    \label{eq:gan_loss} \\
    L_{\text{GP}} &= \lambda \mathbb{E}_{\hat{X}} \left[ \left( \|\nabla_{\hat{X}} D(\hat{X})\|_2 - 1 \right)^2 \right]
    \label{eq:gp_loss} \\
    L_{\text{total}} &= L_{\text{GAN}} + \lambda_{\text{GP}} L_{\text{GP}} + \lambda_{\text{seg}} L_{\text{seg}}
    \label{eq:total_loss}
\end{align}
where Equation~\ref{eq:gan_loss} is the Wasserstein Adversarial Loss, Equation~\ref{eq:gp_loss} is the Gradient Penalty for Stability ($\hat{X}$ is the reference vessel representation) and Equation~\ref{eq:total_loss} is the Structural Consistency via Segmentation Loss. Upon training convergence, the generator produces a vessel structure that not only exhibits realistic contrast properties but also preserves anatomical coherence, ensuring fidelity to real OCTA characteristics.

\subsection{Segmentor}

Rather than utilizing pixel-wise segmentation maps, we directly extract primary vessels from the $(x_{i}^{G}, y_{i}^{G},z_{i}^{G}|r_{i}^{G})$ representation using DFS based graph traversal, which effectively isolates large vessels while filtering out capillaries.

First, we construct a graph $G=(V,E)$, where nodes $V$ represent vessel centerline points and edges $E$ connect adjacent vessel points based on local vessel connectivity. A vessel segment $e_{ij}$ between nodes $v_i$ and $v_j$ (the direction is from $i$ to $j$) is assigned a weight, as shown in Equation~\ref{we}:
\begin{equation}
    w(e_{ij}) = r_i
   \label{we}
\end{equation}
where larger vessels are prioritized in the traversal process. To extract primary vessels, we apply radius thresholding and DFS traversal. For radius filtering, we define the ratio of the smallest main vessel radius to the largest main vessel radius $R_{min}$ and retain only Equation~\ref{rmin}:
\begin{equation}
    V_{\text{main}} = \left\{ (x_i, y_i, z_i | r_i) \mid r_i \geq (R_{\min}*r_{max}) \right\}
   \label{rmin}
\end{equation}

For DFS traversal, it consists of three steps: (1) select the optic disc region as the root node; (2) recursively traverse the largest connected component using DFS; (3) stop when no further vessel segments satisfy the radius constraint. When all three steps finished, construct final extracted vessel structure $G_{main}$. This approach effectively retains only the primary vascular structure, removing small capillaries and noise.

\section{Dataset and Implementation}
\subsection{Dataset}

\subsubsection{OCTA-500}
OCTA-500 \cite{octa-500} focuses on retinal blood vessel research, providing 6mm$\times$6mm (300 images, ID range 10001 to 10300) and 3mm$\times$3mm (200 images, ID range 10301 to 10500) fields of view for wide-field vessel analysis and high-resolution microvascular studies. It includes detailed annotations for arteries, veins, large vessels, capillary networks, and 2D/3D Foveal Avascular Zone, facilitating analysis of retinal vessels structures and pathological changes.

\subsubsection{ROSE}
ROSE \cite{ROSE} is an open-source collection designed for retinal blood vessel segmentation using OCTA images. It consists of two subsets: ROSE-1 includes 117 OCTA images from 39 subjects, covering a 3mm$\times$3mm foveal-centered area with a resolution of 304$\times$304 pixels, providing both centerline-level and pixel-level vessel annotations. ROSE-2 contains 112 OCTA images from 112 eyes, focusing on the superficial vascular complex within a 3mm$\times$3mm area, resized to 840$\times$840 pixels, with centerline-level annotations. 

\begin{figure}[ht]
  \centering
      \begin{subfigure}{0.242\linewidth}
    \centerline{\includegraphics[width=0.8\columnwidth]{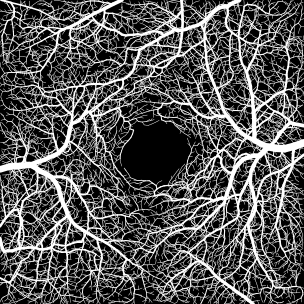}}
    \caption{UTSW Data}
    \label{cid-a}
  \end{subfigure}
    \hfill
  \begin{subfigure}{0.242\linewidth}
\centerline{\includegraphics[width=0.8\columnwidth]{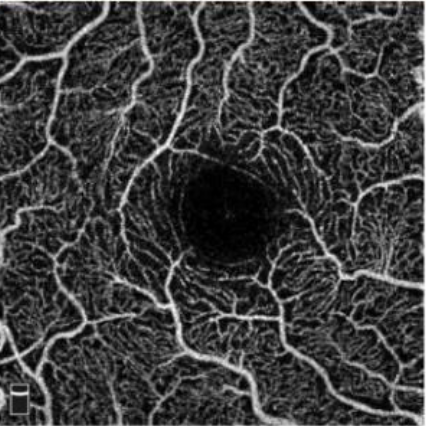}}
    \caption{OCTA-500}
    \label{cid-c}
  \end{subfigure}
  \hfill
  \begin{subfigure}{0.242\linewidth}
\centerline{\includegraphics[width=0.8\columnwidth]{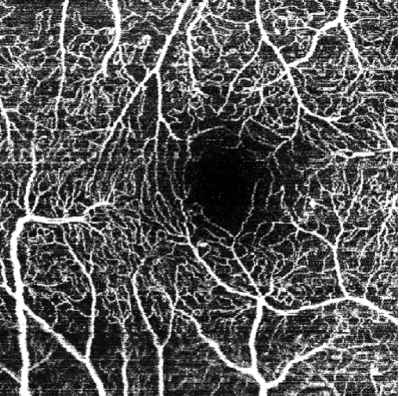}}
    \caption{ROSE}
    \label{cid-b}
  \end{subfigure}
  \caption{Comparison of UTSW with Other Datasets}
  \label{comparison_image_dataset}
\end{figure}

\subsubsection{UTSW}

our partner medical institutions’s institutional review board (IRB) approved this study, which followed the principles of the Declaration of Helsinki. 
Since the study was retrospective, the IRB waived the requirement for informed consent from patients.

As show in Fig.~\ref{comparison_image_dataset}, original images come from the latest Intalight device\footnote{\url{https://intalight.com/}}, which is currently the most advanced equipment. As shown in Fig.~\ref{cid-a}, Intalight images feature clearer vessel boundaries and capillary distributions, enhancing segmentation accuracy. UTSW comprises 550 images, both 2D and 3D, from 250 subjects, ensuring fair representation across gender and race. We collected retinal vascular data for both eyes (OS and OD) of each subject, with a gender split of 48.8\% male and 51.2\% female. Racial distribution reflects local glaucoma clinic demographics: 80.4\% white, 14.6\% Black, with Asians and other minority groups comprising the remainder. Additional data includes age (47.1 $\pm$ 24.8 years) and examination dates from 2023 to 2024. In supervised learning, their minimal pixel variation introduces valuable noise for testing model robustness \cite{octa-500-gen}. For 3D reconstruction and evolution prediction, these high-resolution images aid in vascular modeling and blood flow analysis \cite{octa-500-32d}, making Intalight data ideal for both tasks.

Our team includes two attending glaucoma physicians who are responsible for the detailed labeling of the main retinal blood vessels using the tool Labelme\footnote{\url{https://github.com/wkentaro/labelme}}.

\subsection{Implementation}

\subsubsection{Hyperparameter}

For X-GAN, its generator adopts a ResNet 9-block architecture, while the discriminator utilizes PatchGAN (70$\times$70). The optimizer is Adam with a learning rate of $2\times10^{-4}$ and momentum parameters $\beta_1=0.5$, $\beta_2=0.999$. X-GAN is trained for 50 epochs, with a linear learning rate decay to 0 after 25 epochs, with evaluation performed using 10-fold cross-validation. Data augmentation includes: random rotations ($k\times90^\circ\pm10^\circ$) and flipping.

\subsubsection{Evaluation Matrix}

We use \textit{Intersection over Union} (IoU), also known as the \textit{Jaccard Index}, to measure similarity between the predicted segmentation and the ground truth. This metric ranges from 0 to 1, with 1 indicating perfect segmentation. Additionally, we use the \textit{Dice Coefficient}, another metric for segmentation accuracy that calculates overlap between the prediction and ground truth. All in all, IoU is used to evaluate segmentation accuracy of the main vessels, while the Dice Coefficient assesses accuracy at finer vessel ends relative to the main vessels. For vascular structural enhancement, we use \textit{Structural Similarity Index} (SSIM) and \textit{Mean Squared Error} (MSE). SSIM (range: -1 to 1) assesses similarity based on luminance, contrast, and structure, while MSE quantifies pixel-wise differences, with lower values indicating higher similarity. While SSIM aligns with human perception, MSE focus on absolute pixel differences.

\subsubsection{Experimental Setup}
We selected the following deep learning models as reference benchmarks: U-Net \cite{unet}, U-Net++ \cite{UNet++}, Attention U-Net \cite{aunet}, Mask R-CNN \cite{mask}, YOLOv11-x \cite{v11}, MedSAM \cite{MedSAM}, CauSSL \cite{CauSSL}, UniverSeg \cite{UniverSeg}, S2VNet \cite{S2VNet} and Tyche \cite{Tyche}. These models are leaders in the fields of image segmentation and medical image processing. For their parameter adjustment, we use Optuna\footnote{https://optuna.org/} to optimize their learning green, decay and other hyperparameters. For the training and testing of the model, we use ten cross-validation methods. The evaluation indicators are finally averaged.

\subsubsection{Hardware}
The hardware for training and testing include 2 Tesla V100 GPUs (2 $\times$ 32GB), 64GB of RAM, 8 CPU cores per node, and a total of 6 nodes.

\section{Experimental Result}

\subsection{Comparative Experiment}

\begin{table*}[ht]
\begin{center}
\caption{Comparison X-GAN with Other Models on Different Datasets ($\times$100\%)}
\begin{tabular}{c|cc|cc|cc|cc|cc}
\hline
\multirow{4}{*}{\textbf{Model}}  & \multicolumn{10}{c}{\textbf{Dataset}} \\
\cline{2-11}  &  \multicolumn{4}{c|}{\textbf{UTSW}} & \multicolumn{4}{c|}{\textbf{OCTA-500}} & \multicolumn{2}{c}{\textbf{ROSE}}\\
\cline{2-11}  & \multicolumn{2}{c|}{\textbf{6mm}} & \multicolumn{2}{c|}{\textbf{3mm}} & \multicolumn{2}{c|}{\textbf{6mm}} & \multicolumn{2}{c|}{\textbf{3mm}} & \multicolumn{2}{c}{\textbf{3mm}} \\

\cline{2-11}  &  \textbf{IoU} & \textbf{Dice} & \textbf{IoU} & \textbf{Dice} & \textbf{IoU} & \textbf{Dice} & \textbf{IoU} & \textbf{Dice} & \textbf{IoU} & \textbf{Dice} \\
\hline
U-Net \cite{unet}                 & 95.14 & 97.51 & 95.23 & 97.56 & 96.93 & 98.44 & 98.75 & 99.37 & \underline{97.94} & \underline{98.96} \\ 
U-Net++ \cite{UNet++}             & 95.68 & 97.81 & 96.64 & 98.29 & \underline{97.65} & \underline{98.81} & 97.96 & 98.97 & 97.52 & 98.75 \\ 
Attention U-Net \cite{aunet}      & 95.78 & 97.84 & 96.04 & 97.98 & 97.05 & 98.50 & 98.61 & 99.30 & 97.92 & 98.95 \\
Mask R-CNN \cite{mask}            & 92.35 & 96.02 & 91.35 & 95.48 & 92.06 & 95.87 & 93.50 & 96.64 & 91.23 & 95.41 \\
YOLOv11-x \cite{v11}              & 95.81 & 97.86 & \underline{96.55} & \underline{98.24} & 95.93 & 97.92 & 97.66 & 98.82 & 97.83 & 98.89 \\ 
MedSAM \cite{MedSAM}              & \underline{96.38} & \underline{98.16} & 96.47 & 98.20 & 96.01 & 97.95 & \underline{98.94} & \underline{99.47} & 97.89 & 98.93 \\ 
CauSSL \cite{CauSSL}              & 95.23 & 97.56 & 95.47 & 97.68 & 96.32 & 98.04 & 98.15 & 99.08 & 97.80 & 98.89 \\ 
UniverSeg \cite{UniverSeg}        & 95.30 & 97.58 & 95.52 & 97.70 & 96.58 & 98.17 & 98.42 & 99.22 & 97.07 & 98.01 \\
S2VNet \cite{S2VNet}              & 95.68 & 97.81 & 95.36 & 97.63 & 96.45 & 98.11 & 98.27 & 99.15 & 97.92 & 98.95 \\
Tyche \cite{Tyche}                & 95.80 & 97.85 & 95.19 & 97.53 & 96.28 & 98.02 & 98.11 & 99.06 & 97.76 & 98.87 \\ 
\hline
X-GAN                             & \textbf{99.41} & \textbf{99.71} & \textbf{98.66} & \textbf{99.33} & \textbf{99.21} & \textbf{99.60} & \textbf{99.42} & \textbf{99.71} & \textbf{99.19} & \textbf{99.59} \\ 
\hline
\end{tabular}
\label{er-1}
\end{center}
\end{table*}

As shown in Table~\ref{er-1}, experimental results indicate that while MedSAM, U-Net++, and YOLOv11-x perform competitively, X-GAN consistently outperforms them across resolutions, achieving near-perfect evaluation metrics and excelling in fine-grained retinal vessel segmentation.

As shown in Table~\ref{ss}, comparison between CycleGAN \cite{CycleGAN} and CycleGAN (ours revised version) shows a high structural similarity, with SSIM values ranging from 91.01\% to 96.04\%. CycleGAN (ours) consistently outperforms CycleGAN, achieving higher SSIM and lower MSE across all datasets. The small SSIM difference (less than 4.31\%) indicates that both models generate highly similar images, but CycleGAN (ours) preserves image structure better and reduces pixel-level errors, making it the superior model.

\begin{table*}[ht]
\begin{center}
\caption{Comparison Vessel Structure Refinement with Real OCTA ($\times$100\%)}
\label{ss}
\begin{tabular}{c|cc|cc|cc|cc|cc}
\hline
\multirow{4}{*}{\textbf{Model}}  & \multicolumn{10}{c}{\textbf{Dataset}} \\
\cline{2-11}  &  \multicolumn{4}{c|}{\textbf{UTSW}} & \multicolumn{4}{c|}{\textbf{OCTA-500}} & \multicolumn{2}{c}{\textbf{ROSE}}\\
\cline{2-11}  & \multicolumn{2}{c|}{\textbf{6mm}} & \multicolumn{2}{c|}{\textbf{3mm}} & \multicolumn{2}{c|}{\textbf{6mm}} & \multicolumn{2}{c|}{\textbf{3mm}} & \multicolumn{2}{c}{\textbf{3mm}} \\

\cline{2-11}  &  \textbf{SSIM} & \textbf{MSE} & \textbf{SSIM} & \textbf{MSE} & \textbf{SSIM} & \textbf{MSE} & \textbf{SSIM} & \textbf{MSE} & \textbf{SSIM} & \textbf{MSE} \\
\hline
CycleGAN \cite{CycleGAN}  & 93.75 & 2.36 & 94.32 & 2.39 & 93.56 & 1.62 & 94.58 & 4.46 & 91.01 & 3.83 \\
\hline
CycleGAN (ours) & 95.21 & 1.88 & 95.89 & 1.85 & 94.82 & 1.73 & 96.04 & 3.10 & 95.32 & 2.16 \\
\hline
\end{tabular}
\end{center}
\end{table*}

\noindent\textbf{Human Evaluation}: the medical team verified the corrected images. Luckily, since this study only aimed to segment only the main blood vessels, the small size of the capillaries caused errors in radius generation, which affected the accuracy of the indicators.

\subsection{Ablation Eperiment}

For baseline GAN model, GAN \cite{GAN} and CycleGAN \cite{CycleGAN} are selected. For baseline segmentor, U-Net \cite{unet}, U-Net++ \cite{UNet++}, Attention U-Net \cite{aunet} and MedSAM \cite{MedSAM} are selected. For X-GAN (CycleGAN (ours) + DFS), the dirrerent values of $R_{min}$ are tested to evaluate the model segmentation accuracy. As shown in Table~\ref{er-ab}, since main blood vessels make up a small proportion, the value calculated is also relatively low, aligning with biostatistical properties \cite{kooner2024meeting,bohm2023methods,campbell2017detailed,niemeijer2010automatic}, segmentation accuracy peaks, reinforcing the model’s strong clinical interpretability.

\begin{table}[ht]
\begin{center}
\caption{Ablation Experiment of X-GAN ($\times$100\%)}
\label{er-ab}
\begin{tabular}{l|cc|cc}
\hline
\multirow{4}{*}{\textbf{Model}}  & \multicolumn{4}{c}{\textbf{Dataset}} \\
\cline{2-5}    &\multicolumn{4}{c}{\textbf{UTSW}} \\
\cline{2-5}    & \multicolumn{2}{c|}{\textbf{6mm}} & \multicolumn{2}{c}{\textbf{3mm}}\\
\cmidrule{2-5}    &\textbf{IoU} & \textbf{Dice} & \textbf{IoU} & \textbf{Dice} \\
\hline
GAN \cite{GAN} + U-Net \cite{unet}               & 92.89 & 96.31 & 92.44 & 96.07 \\
GAN \cite{GAN} + U-Net++ \cite{UNet++}           & 93.62 & 96.70 & 92.87 & 96.30 \\
GAN \cite{GAN} + Attention U-Net \cite{aunet}    & 93.56 & 96.67 & 93.12 & 96.44 \\
GAN \cite{GAN} + MedSAM \cite{MedSAM}            & 93.73 & 96.76 & 92.76 & 96.24 \\
GAN \cite{GAN} + DFS ($R_{min}$=0.2)             & 95.81 & 97.86 & 95.24 & 97.56 \\
\hline
CycleGAN \cite{CycleGAN} + U-Net \cite{unet}               & 93.44 & 96.61 & 93.42 & 96.60 \\
CycleGAN \cite{CycleGAN} + U-Net++ \cite{UNet++}           & 93.72 & 96.76 & 93.62 & 96.70 \\
CycleGAN \cite{CycleGAN} + Attention U-Net \cite{aunet}    & 93.79 & 96.80 & 93.45 & 96.61 \\
CycleGAN \cite{CycleGAN} + MedSAM \cite{MedSAM}            & 94.02 & 96.92 & 94.37 & 97.10 \\
CycleGAN \cite{CycleGAN} + DFS ($R_{min}$=0.2)             & 96.23 & 98.08 & 96.42 & 98.18 \\
\hline
CycleGAN (ours) + U-Net \cite{unet}               & 94.56 & 97.20 & 94.52 & 97.18 \\
CycleGAN (ours) + U-Net++ \cite{UNet++}           & 94.98 & 97.43 & 94.36 & 97.10 \\
CycleGAN (ours) + Attention U-Net \cite{aunet}    & 95.06 & 97.47 & 95.12 & 97.50 \\
CycleGAN (ours) + MedSAM \cite{MedSAM}            & 95.33 & 97.61 & 95.21 & 97.55 \\
\hline
X-GAN                             & \textbf{99.41} & \textbf{99.71} & \textbf{98.66} & \textbf{99.33}\\
\hline
\end{tabular}
\end{center}
\end{table}

\begin{figure}[ht]
  \centering
  \begin{tiny}
    \begin{subfigure}{0.40\linewidth}
\centerline{\includegraphics[width=0.95\columnwidth]{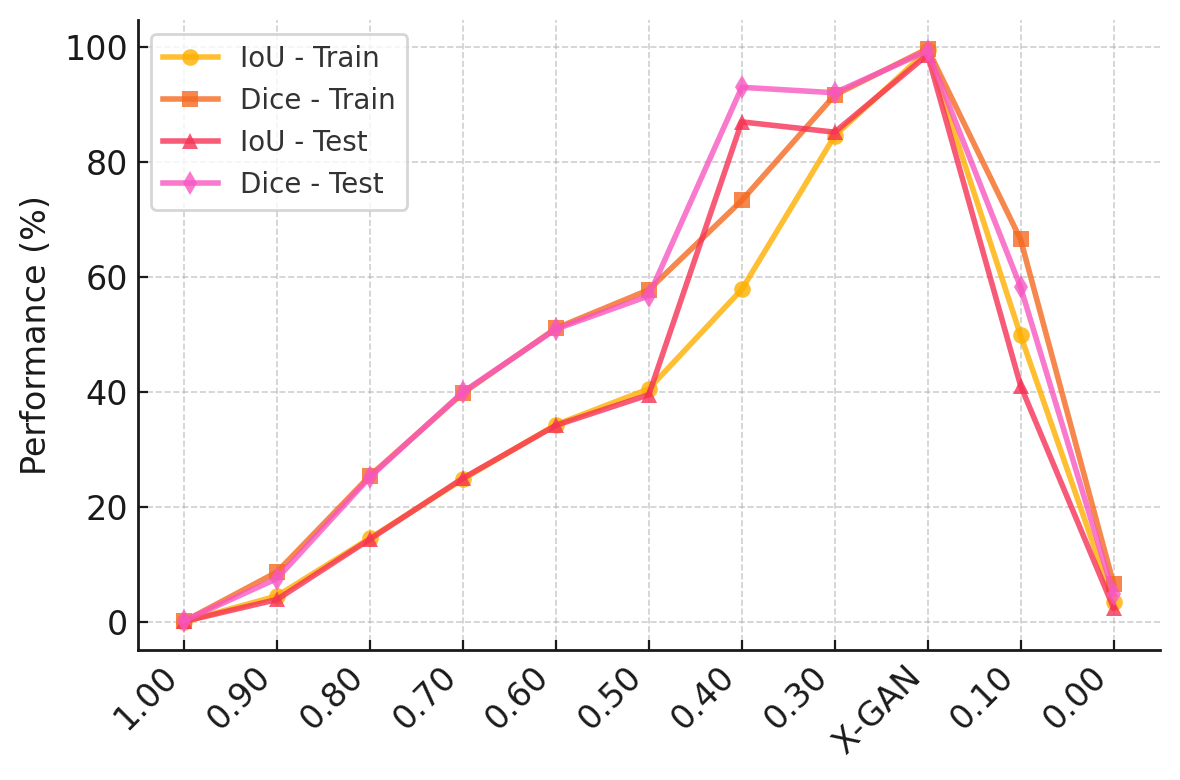}}
    \caption{$R_{min}$}
    \label{ab-1}
  \end{subfigure}
    \hfill
      \begin{subfigure}{0.45\linewidth}
    \centerline{\includegraphics[width=0.95\columnwidth]{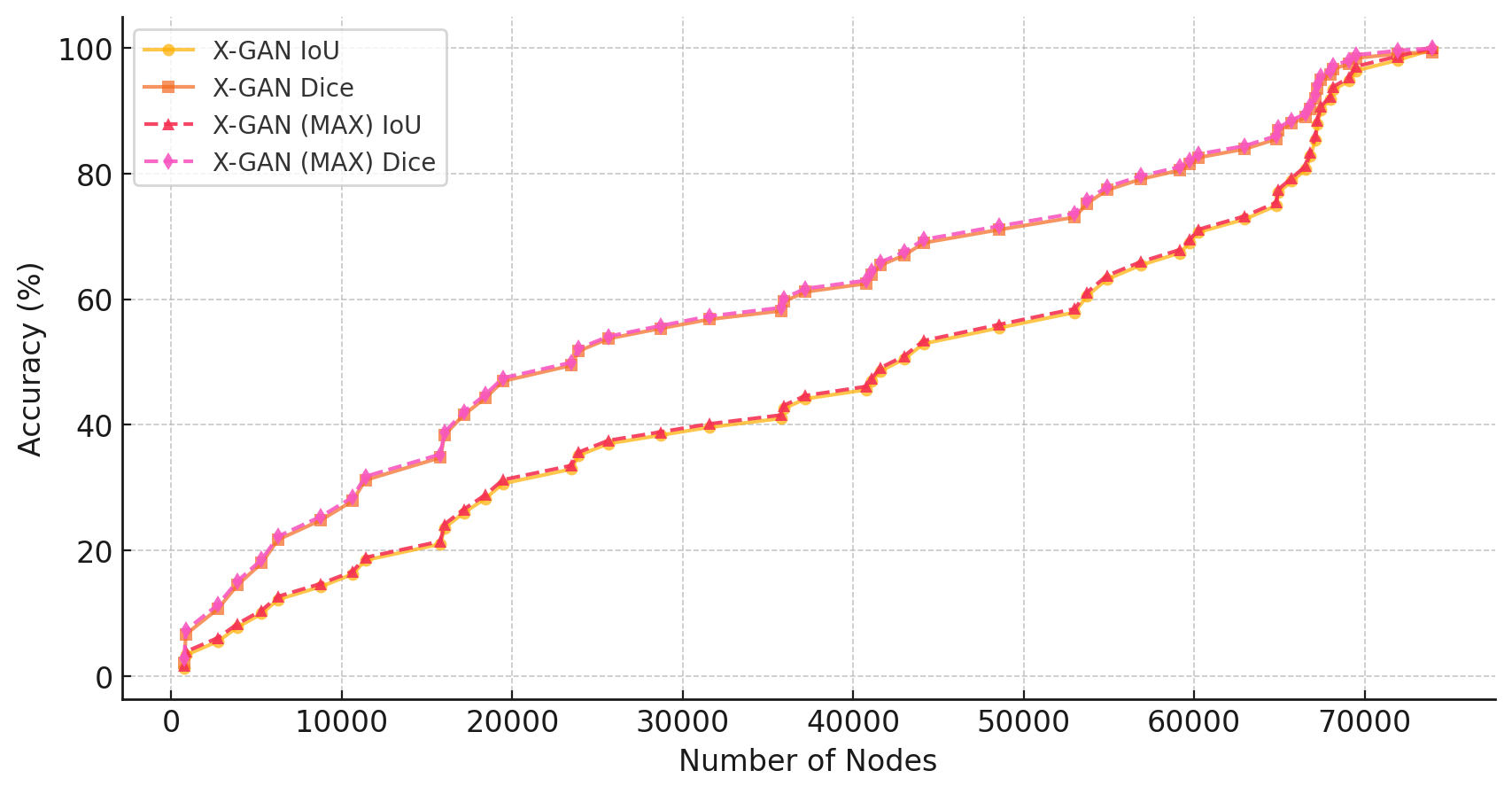}}
    \caption{The Number of Nodes $N$}
    \label{ab-2}
  \end{subfigure}
  \caption{Ablation Experiments of Key Parameters ($\times$100\%)}
  \label{ab}
  \end{tiny}
\end{figure}

As shown in Fig.~\ref{ab-1}, the optimal choice of $R_{min}$ (0.2) enables X-GAN to significantly outperform other models in segmentation, which is so high that it is nearly perfect, approaching 100\%. When $R_{min}=1$, X-GAN segments only the largest vessels, resulting in an extremely low segmentation index as only the starting points meet the threshold. When $R_{min}$ is 0, it is equivalent to no filtering, directly the original image. 

Also, as shown in Fig.~\ref{ab-2}, a greater number of generated nodes $N$ leads to smoother vessel edges, a more complete vascular structure, and improved segmentation accuracy. Given that the R values of X-GAN are relatively low, their practical significance indicates minimal differences in the segmented vessels, as the number of nodes remains similar. 

\subsection{Efficiency Evaluation}

Compared to SLSMs, our segmentor offers these advantages: faster inference (DFS does not need GPU and training, but SLSMs need GPU for training and testing), eliminates the need for large-scale annotations, and ensures full preservation of primary vessel integrity. It features an adjustable capillary filtering parameter $R_{min}$, maintains strong generalization across all vessel OCTA datasets without retraining, and is directly applicable to 3D extensions without additional design. 

\begin{table}[ht]
\begin{center}
\caption{Comparison of Performance Details between DFS (our) and SLSMs}
\label{ccd}
\scalebox{0.95}{
\begin{tabular}{c|c|c|c|c}
\hline
\multirow{2}{*}{\textbf{Model}} &   \multicolumn{4}{c}{\textbf{Details}}   \\
\cline{2-5}& \textbf{Inference Time}     & \textbf{Parameters} & \textbf{FLOPs} & \textbf{GPU Utilization} \\
\hline
\textbf{U-Net} \cite{unet} & 21ms & 31M & 98 & low\\
\textbf{U-Net++} \cite{UNet++} & 62ms & 76M & 304 & high\\
\textbf{Attention U-Net} \cite{aunet}  & 83ms & 62M & 215 & slightly high\\
\textbf{MedSAM} \cite{MedSAM}   & 425ms & 109M & 1250 & extremely high\\
\hline
\textbf{DFS (ours)}    & 0.018ms   & 0M & 0.00028 & -\\
\textbf{DFS (ours, GPU based)}    & $\leq$ 0.01ms   & 0M & 0.00028 & extremely low\\
\hline
\end{tabular}}
\end{center}
\end{table}

Compared to the segmentor, our adjusted DFS offers unmatched advantages, as shown in Table~\ref{ccd}. It requires no separate training or labeling, as DFS is a parameter-efficient algorithm. By eliminating manual annotation, it removes subjective bias from doctors. Additionally, its radius-based segmentation avoids pixel mapping errors in SLSMs, such as interference at vessel edges, inaccuracies at junctions, and mask coverage errors. What is more, the key issue is that both SLSMs and GAN+SLSMs frameworks are susceptible to fitting problems in segmentation tasks. This arises from the training limitations of the SLSMs model and the nature of the data. In contrast, DFS, as an algorithm rather than a model, is unaffected by this issue. 

\subsection{Visualization}
\begin{figure*}[ht]
  \centering
  \begin{subfigure}{1.0\linewidth}
    \centerline{\includegraphics[width=\columnwidth]{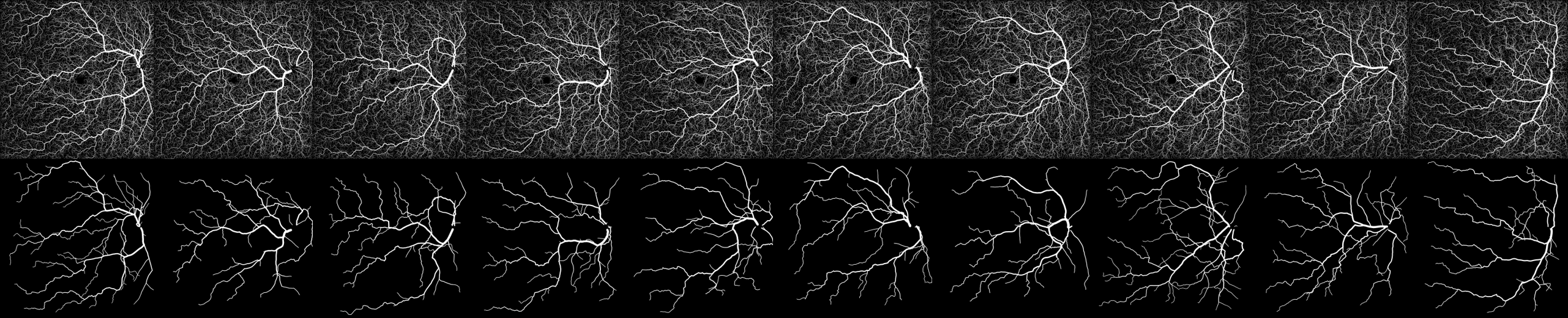}}
    \caption{Segmentation of Main Retinal Vessels (2D)}
    \label{er2d}
  \end{subfigure}
  \hfill
  \begin{subfigure}{0.45\linewidth}
    \centerline{\includegraphics[width=\columnwidth]{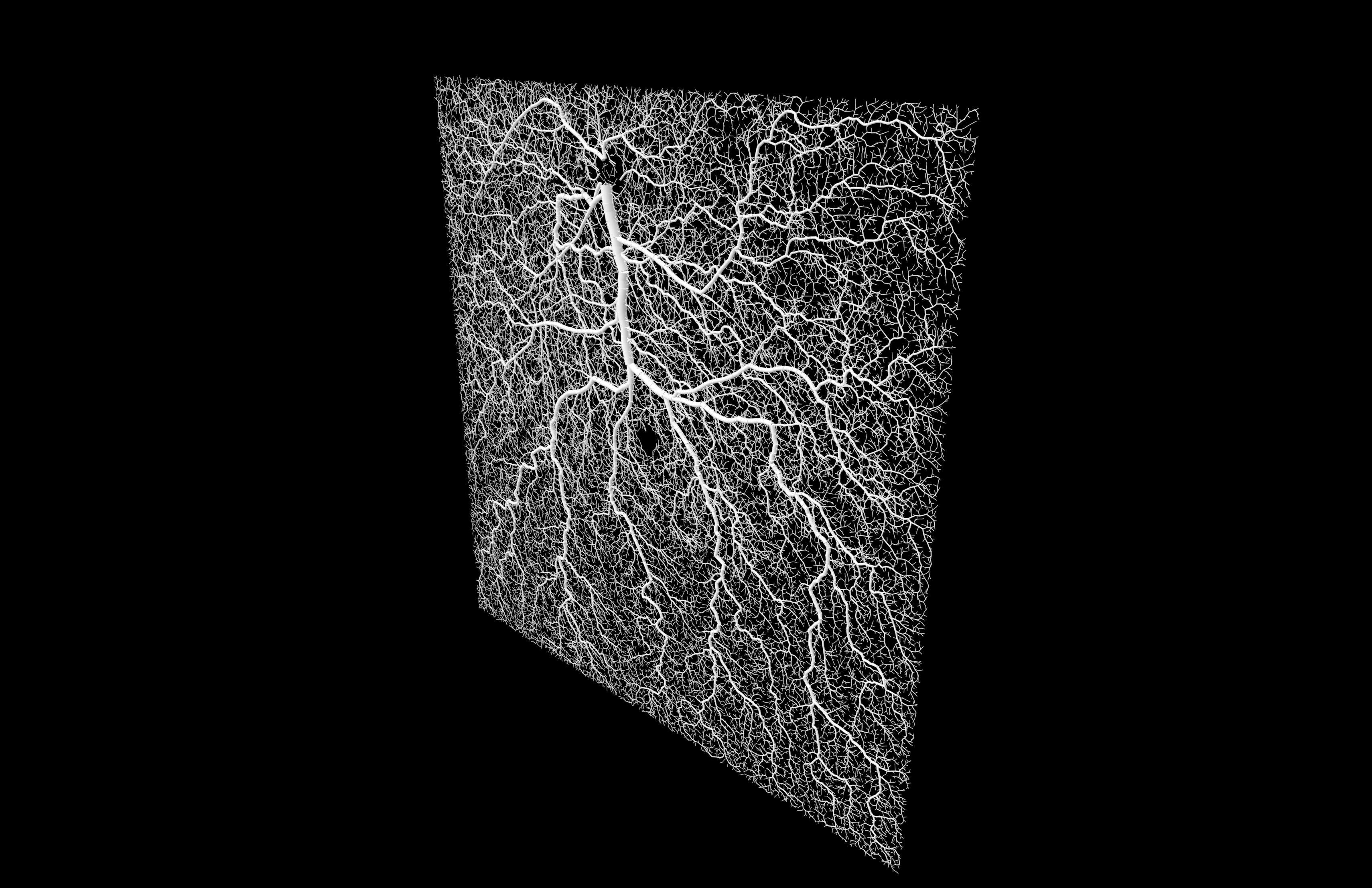}}
    \caption{Original Image (3D)}
    \label{e3d-2}
  \end{subfigure}
      \hfill
  \begin{subfigure}{0.45\linewidth}
    \centerline{\includegraphics[width=\columnwidth]{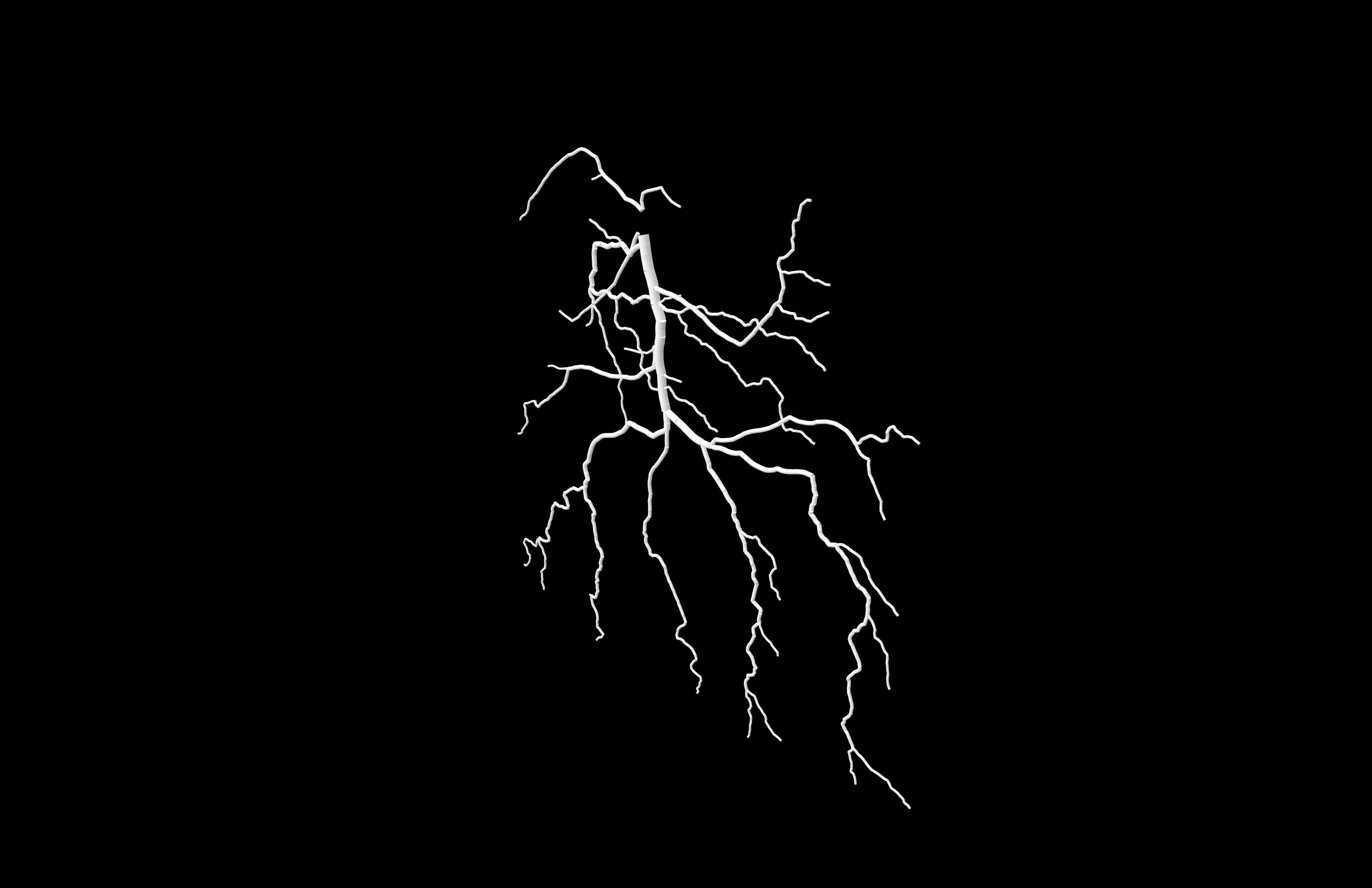}}
    \caption{Segmentation (3D)}
    \label{e3d-4}
  \end{subfigure}

  \caption{Segmentation of X-GAN (2D and 3D)} 
  \label{er-23}
\end{figure*}

As shown in Fig.~\ref{er-23}, our 2D and 3D segmentation results exhibit smooth, well-defined vessel curves, confirming our prior analysis. In Fig.~\ref{er2d}, the segmented vessels are exceptionally smooth, free of pixel artifacts or overflow. DFS effectively filters out vessels below the threshold, eliminating noise and misclassified capillaries. Extending this to 3D, we achieve precise main vessel extraction, with Fig.~\ref{e3d-4} preserving vessel integrity and offering a clearer structure than the original image (Fig.~\ref{e3d-2}).

\subsection{Analysis}

Optimal selection of $R_{min}$ values allows X-GAN to achieve exceptionally high segmentation accuracy, outperforming the second-best model by nearly 3 percentage points, nearly 100\% (Table~\ref{er-1} and Table~\ref{er-ab}). Unlike many SLSMs that have been magically tuned to have very poor interpretability, $R_{min}$ is based on biostatistics of glacucoma \cite{kooner2024meeting,bohm2023methods,campbell2017detailed,niemeijer2010automatic}. We only knew the biological information of the main blood vessel radius and the form of its data expression, and just adjusted X-GAN to achieve amazing segmentation results. Unlike GAN+SLSMs or SLSMs, X-GAN does not require additional data to avoid fitting issues. Its segmentor is efficient, computationally friendly, and free from pixel mapping errors associated with SLSMs segmentation. Studies \cite{convnet,pretrained} have also shown the similar idea: with a deep understanding of data characteristics, the right training strategies, and optimized parameters, even CNNs can outperform transformer-based models.

\section{Conclusion}
In this paper, we propose X-GAN, an unsupervised model for ultra-high-precision segmentation of OCTA retinal main vessels, integrating biostatistical vessel radius properties with a GAN-enhanced DFS algorithm to achieve near-perfect accuracy without labeled data, training, or high-performance GPUs. Extensive experiments demonstrate that X-GAN consistently surpasses SOTA models, achieving nearly 100\% segmentation accuracy, confirming its robustness and generalizability. Our DFS-based segmentation effectively isolates main vessels while filtering out capillary noise, ensuring strong biological interpretability and clinical relevance.

\section{Future Work}

With access to state-of-the-art glaucoma detection equipment, we have curated a high-quality image dataset featuring precisely annotated retinal vascular images. This dataset is well-suited for glaucoma pathology research and computer vision.


\begin{thebibliography}{8}

\bibitem{octa-500}
M. Li, K. Huang, Q. Xu, J. Yang, Y. Zhang, Z. Ji, K. Xie, S. Yuan, Q. Liu, and Q. Chen, ``OCTA-500: a retinal dataset for optical coherence tomography angiography study,'' \textit{Medical Image Analysis}, vol. 93, p. 103092, 2024.

\bibitem{octa-500-32d}
M. Li, Y. Chen, Z. Ji, K. Xie, S. Yuan, Q. Chen, and S. Li, ``Image Projection Network: 3D to 2D Image Segmentation in OCTA Images,'' \textit{IEEE Transactions on Medical Imaging}, vol. 39, no. 11, pp. 3343--3354, 2020.

\bibitem{octa-500-gen}
Y. Zhang, C. Huang, M. Li, S. Xie, K. Xie, Z. Ji, S. Yuan, and Q. Chen, ``Robust layer segmentation against complex retinal abnormalities for en face OCTA generation,'' in \textit{Proc. 23rd Int. Conf. Medical Image Computing and Computer-Assisted Intervention (MICCAI)}, Lima, Peru, Oct. 2020, pp. 647--655.

\bibitem{ROSE}
Y. Ma, H. Hao, J. Xie, H. Fu, J. Zhang, J. Yang, Z. Wang, J. Liu, Y. Zheng, and Y. Zhao, ``ROSE: A Retinal OCT-Angiography Vessel Segmentation Dataset and New Model,'' \textit{IEEE Transactions on Medical Imaging}, vol. 40, no. 3, pp. 928--939, 2021.

\bibitem{dark}
K. He, J. Sun, and X. Tang, ``Single image haze removal using dark channel prior,'' \textit{IEEE Transactions on Pattern Analysis and Machine Intelligence}, vol. 33, no. 12, pp. 2341--2353, 2010.

\bibitem{usm}
Z. Shi, Y. Chen, E. Gavves, P. Mettes, and C. G. M. Snoek, ``Unsharp mask guided filtering,'' \textit{IEEE Transactions on Image Processing}, vol. 30, pp. 7472--7485, 2021.

\bibitem{UNet++}
Z. Zhou, M. M. R. Siddiquee, N. Tajbakhsh, and J. Liang, ``UNet++: Redesigning skip connections to exploit multiscale features in image segmentation,'' \textit{IEEE Transactions on Medical Imaging}, 2019.

\bibitem{MedSAM}
J. Ma, Y. He, F. Li, L. Han, C. You, and B. Wang, ``Segment anything in medical images,'' \textit{Nature Communications}, vol. 15, p. 654, 2024.

\bibitem{mask}
K. He, G. Gkioxari, P. Doll{\'a}r, and R. Girshick, ``Mask R-CNN,'' in \textit{Proc. IEEE Int. Conf. Computer Vision (ICCV)}, 2017, pp. 2961--2969.

\bibitem{unet}
O. Ronneberger, P. Fischer, and T. Brox, ``U-net: Convolutional networks for biomedical image segmentation,'' in \textit{Medical Image Computing and Computer-Assisted Intervention – MICCAI 2015}, Munich, Germany, Oct. 2015, pp. 234--241.

\bibitem{v11}
R. Khanam and M. Hussain, ``Yolov11: An overview of the key architectural enhancements,'' \textit{arXiv preprint arXiv:2410.17725}, 2024.

\bibitem{aunet}
O. Oktay, J. Schlemper, L. Le Folgoc, M. Lee, M. Heinrich, K. Misawa, K. Mori, S. McDonagh, N. Y. Hammerla, B. Kainz, \textit{et al.}, ``Attention U-Net: Learning Where to Look for the Pancreas,'' in \textit{Medical Imaging with Deep Learning}, 2018.

\bibitem{glalstm}
Huang, Cheng, Weizheng Xie, Jian Zhou, Tsengdar Lee, Karanjit Kooner, and Jia Zhang. "GlaLSTM: A Concurrent LSTM Stream Framework for Glaucoma Detection via Biomarker Mining." arXiv preprint arXiv:2408.15555 (2024).

\bibitem{mask-1}
C. Huang, Y. Yu and M. Qi, "Skin Lesion Segmentation Based on Deep Learning," 2020 IEEE 20th International Conference on Communication Technology (ICCT), Nanning, China, 2020, pp. 1360-1364.

\bibitem{CauSSL}
J. Miao, C. Chen, F. Liu, H. Wei, and P.-A. Heng, ``CauSSL: Causality-inspired semi-supervised learning for medical image segmentation,'' in \textit{Proc. IEEE/CVF Int. Conf. Computer Vision (ICCV)}, Oct. 2023, pp. 21426--21437.

\bibitem{UniverSeg}
V. I. Butoi, J. J. G. Ortiz, T. Ma, M. R. Sabuncu, J. Guttag, and A. V. Dalca, ``UniverSeg: Universal medical image segmentation,'' in \textit{Proc. IEEE/CVF Int. Conf. Computer Vision (ICCV)}, Oct. 2023, pp. 21438--21451.

\bibitem{S2VNet}
Y. Ding, L. Li, W. Wang, and Y. Yang, ``Clustering propagation for universal medical image segmentation,'' in \textit{Proc. IEEE/CVF Conf. Computer Vision and Pattern Recognition (CVPR)}, Jun. 2024, pp. 3357--3369.

\bibitem{glaboost}
Huang, Cheng, Weizheng Xie, Karanjit Kooner, Tsengdar Lee, Jui-Kai Wang, and Jia Zhang. "GlaBoost: A multimodal Structured Framework for Glaucoma Risk Stratification." arXiv preprint arXiv:2508.03750 (2025).




\bibitem{Tyche}
M. Rakic, H. E. Wong, J. J. G. Ortiz, B. A. Cimini, J. V. Guttag, and A. V. Dalca, ``Tyche: Stochastic in-context learning for medical image segmentation,'' in \textit{Proc. IEEE/CVF Conf. Computer Vision and Pattern Recognition (CVPR)}, Jun. 2024, pp. 11159--11173.

\bibitem{GAN-Unet}
L. Kreitner, J. C. Paetzold, N. Rauch, C. Chen, A. M. Hagag, A. E. Fayed, S. Sivaprasad, S. Rausch, J. Weichsel, B. H. Menze, M. Harders, B. Knier, D. Rueckert, and M. J. Menten, ``Synthetic optical coherence tomography angiographs for detailed retinal vessel segmentation without human annotations,'' \textit{IEEE Transactions on Medical Imaging}, vol. 43, no. 6, pp. 2061--2073, 2024.


\bibitem{gan-d-1}
J. Kim, C. Oh, H. Do, S. Kim, and K. Sohn, ``Diffusion-driven GAN inversion for multi-modal face image generation,'' in \textit{Proc. IEEE/CVF Conf. Computer Vision and Pattern Recognition (CVPR)}, Jun. 2024, pp. 10403--10412.

\bibitem{GU-2}
C. Wu, Y. Zou, and Z. Yang, ``U-GAN: Generative adversarial networks with U-Net for retinal vessel segmentation,'' in \textit{Proc. 14th Int. Conf. Computer Science \& Education (ICCSE)}, 2019, pp. 642--646.

\bibitem{DO-GAN}
A. P. P. Aung, X. Wang, R. Yu, B. An, S. Jayavelu, and X. Li, ``DO-GAN: A double oracle framework for generative adversarial networks,'' in \textit{Proc. IEEE/CVF Conf. Computer Vision and Pattern Recognition (CVPR)}, Jun. 2022, pp. 11275--11284.

\bibitem{DCP-GAN}
J. Chung, S. Hyun, S.-H. Shim, and J.-P. Heo, ``Diversity-aware channel pruning for StyleGAN compression,'' in \textit{Proc. IEEE/CVF Conf. Computer Vision and Pattern Recognition (CVPR)}, Jun. 2024, pp. 7902--7911.

\bibitem{vit}
A. Dosovitskiy, L. Beyer, A. Kolesnikov, D. Weissenborn, X. Zhai, T. Unterthiner, M. Dehghani, M. Minderer, G. Heigold, S. Gelly, \textit{et al.}, ``An image is worth 16x16 words: Transformers for image recognition at scale,'' \textit{arXiv preprint arXiv:2010.11929}, 2020.

\bibitem{CNN-GAN}
A. P. Behera, S. Prakash, S. Khanna, S. Nigam, and S. Verma, ``CNN-based metrics for performance evaluation of generative adversarial networks,'' \textit{IEEE Transactions on Artificial Intelligence}, vol. 5, no. 10, pp. 5040--5049, 2024.

\bibitem{GAN-m1}
Z. Zhou, Y. Wang, Y. Guo, X. Jiang, and Y. Qi, ``Ultrafast plane wave imaging with line-scan-quality using an ultrasound-transfer generative adversarial network,'' \textit{IEEE Journal of Biomedical and Health Informatics}, vol. 24, no. 4, pp. 943--956, 2020.

\bibitem{ct-gan}
C. You, G. Li, Y. Zhang, X. Zhang, H. Shan, M. Li, S. Ju, Z. Zhao, Z. Zhang, W. Cong, M. W. Vannier, P. K. Saha, E. A. Hoffman, and G. Wang, ``CT super-resolution GAN constrained by the identical, residual, and cycle learning ensemble (GAN-CIRCLE),'' \textit{IEEE Transactions on Medical Imaging}, vol. 39, no. 1, pp. 188--203, 2020.

\bibitem{TR-GAN}
C.-C. Fan, L. Peng, T. Wang, H. Yang, X.-H. Zhou, Z.-L. Ni, G. Wang, S. Chen, Y.-J. Zhou, and Z.-G. Hou, ``TR-GAN: Multi-session future MRI prediction with temporal recurrent generative adversarial network,'' \textit{IEEE Transactions on Medical Imaging}, vol. 41, no. 8, pp. 1925--1937, 2022.

\bibitem{skin1}
C. Huang, A. Yu, Y. Wang and H. He, "Skin Lesion Segmentation Based on Mask R-CNN," 2020 International Conference on Virtual Reality and Visualization (ICVRV), Recife, Brazil, 2020, pp. 63-67.

\bibitem{GAN-cance}
J. Lee and R. M. Nishikawa, ``Identifying women with mammographically-occult breast cancer leveraging GAN-simulated mammograms,'' \textit{IEEE Transactions on Medical Imaging}, vol. 41, no. 1, pp. 225--236, 2022.

\bibitem{convnet}
Z. Liu, H. Mao, C.-Y. Wu, C. Feichtenhofer, T. Darrell, and S. Xie, ``A convnet for the 2020s,'' in \textit{Proc. IEEE/CVF Conf. Computer Vision and Pattern Recognition (CVPR)}, 2022, pp. 11976--11986.

\bibitem{pretrained}
Y. Tay, M. Dehghani, J. P. Gupta, V. Aribandi, D. Bahri, Z. Qin, and D. Metzler, ``Are pretrained convolutions better than pretrained transformers?,'' in \textit{Proc. 59th Annu. Meeting Assoc. Comput. Linguistics and 11th Int. Joint Conf. Natural Language Processing (Vol. 1: Long Papers)}, 2021, pp. 4349--4359.


\bibitem{kooner2024meeting}
K. S. Kooner, D. M. Choo, and P. Mekala, ``Meeting challenges in the diagnosis and treatment of glaucoma,'' \textit{Bioengineering}, vol. 12, no. 1, p. 6, 2024.

\bibitem{bohm2023methods}
E. W. Böhm, N. Pfeiffer, F. M. Wagner, and A. Gericke, ``Methods to measure blood flow and vascular reactivity in the retina,'' \textit{Frontiers in Medicine}, vol. 9, p. 1069449, 2023.

\bibitem{campbell2017detailed}
J. P. Campbell, M. Zhang, T. S. Hwang, S. T. Bailey, D. J. Wilson, Y. Jia, and D. Huang, ``Detailed vascular anatomy of the human retina by projection-resolved optical coherence tomography angiography,'' \textit{Scientific Reports}, vol. 7, p. 42201, 2017.

\bibitem{niemeijer2010automatic}
M. Niemeijer, B. van Ginneken, and M. D. Abr{\`a}moff, ``Automatic determination of the artery-vein ratio in retinal images,'' in \textit{Medical Imaging 2010: Computer-Aided Diagnosis}, vol. 7624, pp. 143--152, 2010.

\bibitem{GAN}
I. Goodfellow, J. Pouget-Abadie, M. Mirza, B. Xu, D. Warde-Farley, S. Ozair, A. Courville, and Y. Bengio, ``Generative adversarial nets,'' \textit{Advances in Neural Information Processing Systems}, vol. 27, 2014.

\bibitem{CycleGAN}
J.-Y. Zhu, T. Park, P. Isola, and A. A. Efros, ``Unpaired image-to-image translation using cycle-consistent adversarial networks,'' in \textit{Proc. IEEE Int. Conf. Computer Vision}, 2017, pp. 2223--2232.

\bibitem{fairseg}
Y. Tian, M. Shi, Y. Luo, A. Kouhana, T. Elze, and M. Wang, ``FairSeg: A large-scale medical image segmentation dataset for fairness learning using segment anything model with fair error-bound scaling,'' in \textit{Proc. Int. Conf. Learning Representations (ICLR)}, 2023.

\bibitem{fairclip}
Y. Luo, M. Shi, M. O. Khan, M. M. Afzal, H. Huang, S. Yuan, Y. Tian, L. Song, A. Kouhana, T. Elze, \textit{et al.}, ``Fairclip: Harnessing fairness in vision-language learning,'' in \textit{Proc. IEEE/CVF Conf. Computer Vision and Pattern Recognition (CVPR)}, 2024, pp. 12289--12301.

\bibitem{fairdomain}
Y. Tian, C. Wen, M. Shi, M. M. Afzal, H. Huang, M. O. Khan, Y. Luo, Y. Fang, and M. Wang, ``FairDomain: Achieving fairness in cross-domain medical image segmentation and classification,'' \textit{arXiv preprint arXiv:2407.08813}, 2024.

\bibitem{Fairness}
Y. Luo, Y. Tian, M. Shi, L. R. Pasquale, L. Q. Shen, N. Zebardast, T. Elze, and M. Wang, ``Harvard Glaucoma Fairness: A retinal nerve disease dataset for fairness learning and fair identity normalization,'' \textit{IEEE Transactions on Medical Imaging}, vol. 43, no. 7, pp. 2623--2633, 2024.

\bibitem{GDP}
Y. Luo, M. Shi, Y. Tian, T. Elze, and M. Wang, ``Harvard Glaucoma Detection and Progression: A multimodal multitask dataset and generalization-reinforced semi-supervised learning,'' in \textit{Proc. IEEE/CVF Int. Conf. Computer Vision (ICCV)}, 2023, pp. 20471--20482.

\bibitem{mask2}
C. Huang, A. Yu, Y. Wang and H. He, "Skin Lesion Segmentation Based on Mask R-CNN," 2020 International Conference on Virtual Reality and Visualization (ICVRV), Recife, Brazil, 2020, pp. 63-67.


\bibitem{FairVision}
Y. Luo, M. O. Khan, Y. Tian, M. Shi, Z. Dou, T. Elze, Y. Fang, and M. Wang, ``FairVision: Equitable deep learning for eye disease screening via fair identity scaling,'' \textit{arXiv preprint arXiv:2310.02492}, 2024.

\bibitem{kooner2022glaucoma}
K. S. Kooner, A. Angirekula, A. H. Treacher, G. Al-Humimat, M. F. Marzban, A. Chen, R. Pradhan, N. Tunga, C. Wang, P. Ahuja, \textit{et al.}, ``Glaucoma diagnosis through the integration of optical coherence tomography/angiography and machine learning diagnostic models,'' \textit{Clinical Ophthalmology (Auckland, NZ)}, vol. 16, p. 2685, 2022.

\bibitem{lag1}
L. Li, M. Xu, X. Wang, L. Jiang, and H. Liu, ``Attention-based glaucoma detection: A large-scale database and CNN model,'' in \textit{Proc. IEEE Conf. Computer Vision and Pattern Recognition (CVPR)}, Jun. 2019.

\bibitem{cnn-2}
Z. Li, Y. He, S. Keel, W. Meng, R. T. Chang, and M. He, ``Efficacy of a deep learning system for detecting glaucomatous optic neuropathy based on color fundus photographs,'' \textit{Ophthalmology}, vol. 125, no. 8, pp. 1199--1206, 2018.

\bibitem{cnn-1}
A. Li, J. Cheng, D. W. K. Wong, and J. Liu, ``Integrating holistic and local deep features for glaucoma classification,'' in \textit{Proc. 38th Annu. Int. Conf. IEEE Engineering in Medicine and Biology Society (EMBC)}, 2016, pp. 1328--1331.

\bibitem{unet-gla}
J. Kim, L. Tran, E. Y. Chew, and S. Antani, ``Optic disc and cup segmentation for glaucoma characterization using deep learning,'' in \textit{Proc. IEEE 32nd Int. Symp. Computer-Based Medical Systems (CBMS)}, 2019, pp. 489--494.

\bibitem{traditional}
Y. Giarratano, E. Bianchi, C. Gray, A. Morris, T. MacGillivray, B. Dhillon, and M. O. Bernabeu, ``Automated segmentation of optical coherence tomography angiography images: Benchmark data and clinically relevant metrics,'' \textit{Translational Vision Science \& Technology}, vol. 9, no. 13, pp. 5--5, 2020.

\bibitem{meiburger2021automatic}
K. M. Meiburger, M. Salvi, G. Rotunno, W. Drexler, and M. Liu, ``Automatic segmentation and classification methods using optical coherence tomography angiography (OCTA): A review and handbook,'' \textit{Applied Sciences}, vol. 11, no. 20, p. 9734, 2021.

\bibitem{sca}
N. Rauch and M. Harders, ``Interactive synthesis of 3D geometries of blood vessels,'' in \textit{Eurographics 2021 - Short Papers}, H. Theisel and M. Wimmer, Eds. The Eurographics Association, 2021. DOI: 10.2312/egs.20211012.

\bibitem{gai}
X. Xing, F. Shi, J. Huang, Y. Wu, Y. Nan, S. Zhang, Y. Fang, M. Roberts, C.-B. Schönlieb, J. Del Ser, \textit{et al.}, ``When AI eats itself: On the caveats of data pollution in the era of generative AI,'' \textit{arXiv preprint arXiv:2405.09597}, 2024.

\bibitem{tarjan1972dfs}
R. E. Tarjan, ``Depth-first search and linear graph algorithms,'' \textit{SIAM Journal on Computing}, vol. 1, no. 2, pp. 146--160, 1972.

\bibitem{dataset-DRIVE}
J. Staal, M. D. Abramoff, M. Niemeijer, M. A. Viergever, and B. van Ginneken, ``Ridge-based vessel segmentation in color images of the retina,'' \textit{IEEE Transactions on Medical Imaging}, vol. 23, no. 4, pp. 501--509, 2004.

\bibitem{dataset-ORIGA}
Z. Zhang, F. S. Yin, J. Liu, W. K. Wong, N. M. Tan, B. H. Lee, J. Cheng, and T. Y. Wong, ``Origa-light: An online retinal fundus image database for glaucoma analysis and research,'' in \textit{Proc. Annu. Int. Conf. IEEE Eng. Med. Biol.}, 2010, pp. 3065--3068.

\bibitem{dataset-Drishti-GS}
J. Sivaswamy, S. R. Krishnadas, G. D. Joshi, M. Jain, and A. U. S. Tabish, ``Drishti-GS: Retinal image dataset for optic nerve head(ONH) segmentation,'' in \textit{Proc. IEEE Int. Symp. Biomedical Imaging (ISBI)}, 2014, pp. 53--56.

\bibitem{sca-1}
L. Lin, L. Peng, H. He, P. Cheng, J. Wu, K. K. Y. Wong, and X. Tang, ``YOLOCurvSeg: You only label one noisy skeleton for vessel-style curvilinear structure segmentation,'' \textit{Medical Image Analysis}, vol. 90, p. 102937, 2023.

\bibitem{sca-2}
J. J. Lee, B. Li, S. Beery, J. Huang, S. Fei, R. A. Yeh, and B. Benes, ``Tree-D fusion: Simulation-ready tree dataset from single images with diffusion priors,'' in \textit{Proc. European Conf. Computer Vision (ECCV)}, 2024, pp. 439--460.




\end{thebibliography}
\end{document}